\documentclass[prl,10pt,twocolumn,showpacs]{revtex4}
\usepackage[dvips]{epsfig}
\usepackage{float}
\usepackage{amsfonts}
\usepackage{amsmath}
\usepackage{amssymb}
\usepackage{enumerate}         
\usepackage{pslatex}

\begin{document}

\title{Integration of fiber coupled high-$Q$ SiN$_{x}$ microdisks with atom chips}

\author{Paul E. Barclay}
\email{pbarclay@caltech.edu}
\author{Kartik Srinivasan}
\author{Oskar Painter}
\affiliation{Thomas J. Watson, Sr.\ Laboratory of Applied Physics, California Institute of Technology, Pasadena, CA 91125, USA.}
\author{Benjamin Lev}
\author{Hideo Mabuchi}
\affiliation{Norman Bridge Laboratory of Physics, California Institute of Technology, Pasadena, CA 91125, USA.}
\date{\today}

\begin{abstract}
  Micron scale silicon nitride (SiN$_x$) microdisk optical resonators are demonstrated with $Q = 3.6 \times 10^6$ and an
  effective mode volume of $15 (\lambda / n)^3$ at near visible wavelengths.  A hydrofluoric acid wet etch provides
  sensitive tuning of the microdisk resonances, and robust mounting of a fiber taper provides efficient fiber optic
  coupling to the microdisks while allowing unfettered optical access for laser cooling and trapping of atoms.
  Measurements indicate that cesium adsorption on the SiN$_x$ surfaces significantly red-detunes the microdisk
  resonances.  A technique for parallel integration of multiple (10) microdisks with a single fiber taper is also
  demonstrated.
\end{abstract}

\maketitle

\noindent 

Atom chip technology\cite{ref:Reichel,ref:Folman} has rapidly evolved over the last decade as a valuable tool in
experiments involving the cooling, trapping, and transport of ultra-cold neutral atom clouds.  During the same period
there has been significant advancement in microphotonic systems\cite{ref:Vahala4} for the guiding and trapping of light
in small volumes, with demonstrations of photonic crystal nanocavities capable of efficiently trapping light within a
cubic wavelength\cite{ref:Noda5} and chip-based silica microtoroid resonators\cite{ref:Vahala3} with photon lifetimes
well over $10^8$ optical cycles.  Poised to significantly benefit from these developments is the field of cavity quantum
electrodynamics (cavity QED)\cite{ref:Kimble1}, in which strong interactions of atoms with light inside a resonant
cavity can be used to aid in quantum information processing and in the communication and distribution of quantum
information within a quantum network\cite{ref:Duan1}.  Integration of atomic and microphotonic
chips\cite{ref:Rosenblit,ref:Lev,ref:Haase1} offers several advancements to the current state-of-the-art Fabry-Perot
cavity QED systems\cite{ref:McKeever1}, most notably a scalable platform for locally controlling multiple quantum bits
and an increased bandwidth of operation.  In this Letter we demonstrate the suitability of silicon nitride (SiN$_x$) for
high-$Q$, small mode volume microcavities resonant at near-visible wavelengths necessary for cavity QED with alkali
atoms, and describe a robust mounting technique which enables the integration of a permanently fiber-coupled microdisk
resonator with a magnetostatic atom chip.

\begin{figure}[ht]
\begin{center}
  \epsfig{figure=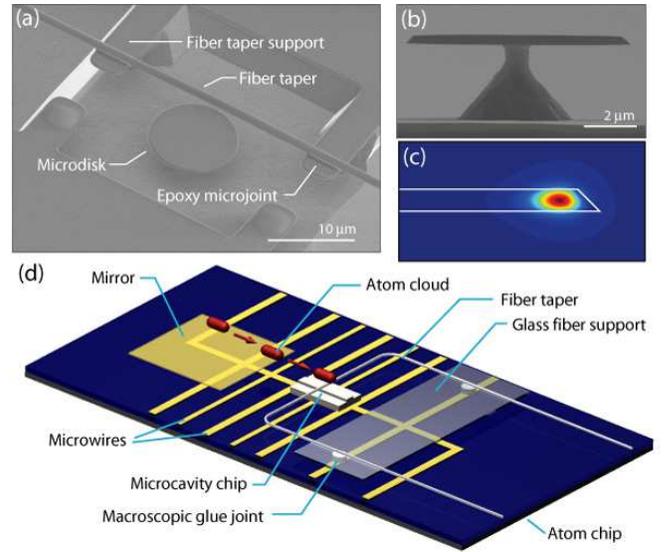, width=1\linewidth}
\caption{(a) Scanning electron microscope (SEM) image of a SiN$_x$ cavity coupled to an optical fiber taper.  The fiber
taper is permanently aligned a few hundred nanometers from the microdisk circumference with epoxy microjoints to SiN$_x$
supports.  (b) Side-view SEM image of a 9 $\mu$m diameter mircodisk. (c) FEM calculated field distribution ($|E|^2$) of
a $m=50$, $p=1$, TE-like mode of the microdisk in (b).  (d) Schematic of the integrated hybrid atom-cavity chip.
}
\label{fig:SEM}
\end{center}
\end{figure}

In addition to the obvious benefits of the fabrication maturity of the silicon(Si)-silicon oxide(SiO$_x$) materials
system, recent work has shown that high quality atom chips can be created from thermally evaporated gold metal wires on
thin oxide coated Si wafers\cite{ref:Groth}.  Integration with a SiN$_x$ optical layer provides a path towards a
monolithic atom-cavity chip with integrated atomic and photonic functionality.  Indeed, owing to its moderately high
index of refraction ($n \sim 2.0$-$2.5$) and large transparency window ($6$ $\mu$m $> \lambda > 300$
nm)\cite{ref:Parsons,ref:Inukai}, SiN$_x$ is an excellent material for the on-chip guiding and localization of light.
The high refractive index of SiN$_x$ makes possible the creation of a variety of wavelength scale microcavity geometries
such as whispering-gallery \cite{ref:Little,ref:Barwicz} or planar photonic crystal structures\cite{ref:Netti}, with a
small intrinsic radiation loss.  Combined with a lower index SiO$_x$ cladding, waveguiding in a SiN$_x$ layer can be
used to distribute light within a planar microphotonic circuit suitable for high-density integration.  The low
absorption loss across the visible and near-IR wavelengths, on the other hand, allows SiN$_x$ to be used with a diverse
set of atomic and atomic-like (colloidal quantum dots, color centers, etc.) species.  Beyond the particular focus of
this work on cavity QED experiments with cold alkali atoms, high-$Q$ SiN$_x$ microcavities are also well suited to
experiments involving moderate refractive index environments, such as sensitive detection of analytes contained in a
fluid solution\cite{ref:Arnold} or absorbed into a low index polymer cladding\cite{ref:Ksendzov2}.

The SiN$_x$ microdisk resonators in this work were fabricated from a commercially available Si wafer with a $250$ nm
thick stoichiometric SiN$_x$ ($n = 2.0$) layer grown on the surface by low pressure chemical vapor deposition (LPCVD).
Fabrication of the microdisk resonators involved several steps, beginning with the creation of a highly circular
electron beam resist etch mask through electron beam lithography and a resist reflow\cite{ref:Borselli2}. A
C$_4$F$_8$/SF$_6$ plasma dry etch was optimized to transfer the resist etch mask into the SiN$_x$ layer as smoothly as
possible.  This was followed by a potassium hydroxide wet etch to selectively remove the underlying $\langle 100\rangle$
Si substrate until the SiN$_x$ microdisks were supported by a small micron diameter silicon pillar.  A final cleaning
step to remove organic materials from the disk surface was performed using a H$_2$SO$_4$:H$_2$O$_2$ wet etch.  A SEM
image of a fully processed microdisk is shown in Fig 1(a,b).

\begin{figure}[h]
\begin{center}
  \epsfig{figure=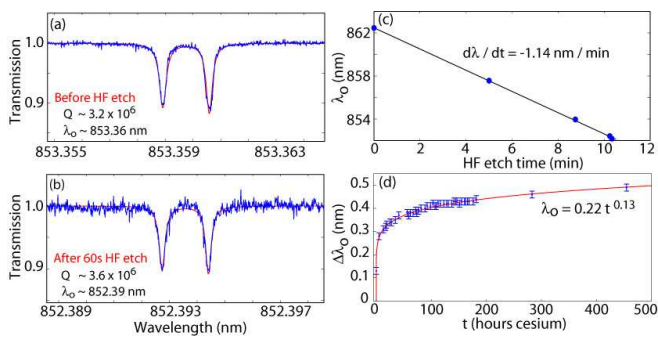,
    width=1.0\linewidth}
\caption{Wavelength scan of the fiber taper transmission for a $p=1$ TE-like mode of a $9$ $\mu$m diameter SiN$_x$
microdisk prior to (a) and (b) and after HF wet etch tuning of the resonance wavelength.  (c) Resonance wavelength
($\lambda_{o}$) of the $p=1$ TE-like mode of an $11$ $\mu$m diameter microdisk as a function of HF etch time.  (d) Shift
in $\lambda_o$ as a function of time exposed to cesium vapor (partial pressure $\sim 10^{-9}$ Torr) in a UHV chamber.}
\label{fig:HF_tuning}
\end{center}
\end{figure}

The optical modes of the fabricated microdisks were efficiently excited via an optical fiber taper
waveguide\cite{ref:Knight,ref:Spillane2}.  A swept wavelength source covering the $840$-$856$ nm wavelength band was
coupled into the fiber taper waveguide and used to measure the fine transmission spectra of the microdisk resonators at
wavelengths close to the D2 transition of cesium (Cs).  Details of the fiber taper measurement set-up can be found in
Ref. \onlinecite{ref:Borselli2}.  Figure 2(a) shows a typical measured wavelength scan of the lowest
radial order ($p=1$) TE-like mode of a $9$ $\mu$m diameter SiN$_x$ microdisk.  The resonance has an intrinsic linewidth
of $0.26$ pm, corresponding to a quality factor $Q = 3.6 \times 10^6$.  The doublet structure in the transmission
spectra is due to mode-coupling between the clockwise and counter-clockwise modes of the disk due to surface roughness
induced backscattering\cite{ref:Kippenberg}.  In addition to this high-$Q$ mode, the $9$ $\mu$m microdisks also support
a lower $Q$ higher-order radial mode in the $850$ nm wavelength band.  The free spectral range between modes of the same
radial order but different azimuthal number ($m$) was measured to be $5.44$ THz (13 nm), resulting in a finesse of
$\mathcal{F} = 5 \times 10^4$ for the $p=1$, TE-like modes.  Tests of less surface sensitive larger diameter microdisks
showed reduced doublet splitting but no reduction in linewidth, indicating that $Q$ is most likely limited by material
absorption and not surface roughness\cite{ref:Borselli2}.  This bodes well for utilizing these SiN$_x$ microdisks at
even shorter wavelengths, such as the $780$ nm D2 transition of rubidium.

Finite-element method (FEM) simulations\cite{ref:Borselli2} (Fig.\ 1(c)) show that the effective optical
mode volume, defined by $V_{\text{eff}} =\int{n^2(\mathbf{r})E^2(\mathbf{r})
  d\mathbf{r}}/|n^2(\mathbf{r})E^2(\mathbf{r})|_{\text{max}}$, is as small as $15$ $(\lambda/n)^3$ for the $9$ $\mu$m
diameter mirodisks.  The corresponding parameters of cavity QED, the Cs-photon coherent coupling rate ($g$), photon
field decay rate ($\kappa$), and Cs transverse decay rate ($\gamma_\perp$) are
$\left[g,\kappa,\gamma_{\perp}\right]/2\pi = \left[2.4, 0.05, 0.003\right]$ GHz for an atom at the mircodisk surface,
indicating that these cavities are capable of operating well within the regime of strong coupling\cite{ref:Kimble1}.
For an atom displaced 100 nm from the surface, $g / 2 \pi$ drops to $0.9$ GHz.  FEM simulations show that for microdisks
of diameter below $9$ $\mu$m the intrinsic radiation $Q$ drops rapidly below the $10^8$ level.

Using the above fabrication procedure the resonance wavelength ($\lambda_o$) of the microdisk modes could be positioned
with an accuracy of $\pm 0.5$ nm.  In order to finely tune $\lambda_o$ into alignment with the D2 atomic Cs transition a
series of timed etches in 20:1 diluted 49\% HF solution was employed.  By slowly etching the LPCVD SiN$_x$ the resonance
wavelength of the high-$Q$ disk modes was shown to blue shift at a rate of $1.1$ nm$/$min (Fig.\ 
2(c)).  With this technique the cavity resonance could be positioned with an accuracy of $\pm 0.05$ nm
without degrading the $Q$ factor (Fig. \ 2(b)).  Further fine tuning can be accomplished by heating
and cooling of the sample; a temperature dependence of $d\lambda_o / dT \sim 0.012$ $\text{nm} / ^\text{o} \text{C}$ was
measured for the $p=1$, TE-like microdisk modes.

After initial device characterization and tuning of $\lambda_o$, the fiber taper and microdisk chip were integrated with
an atom chip consisting of a sapphire substrate with electroplated gold microwires underneath a top evaporated gold
mirror layer\cite{ref:Lev2,ref:Lev}.  A brief outline of the integration procedure follows.  The 3x3x0.3 mm Si
microcavity chip is aligned and bonded to the desired location on the top surface of the atom chip using polymethyl
methacrylate.  The fiber taper is supported in a self-tensioning ``U'' configuration by a glass coverslip ($\sim 200$
$\mu$m thick) as illustrated in Fig.\ 1(d).  The taper is aligned with the microdisk using DC motor stages with 50 nm
encoder resolution.  Adjustment in the lateral gap between the taper and the microdisk is used to tune the level of
cavity loading; owing to the excellent phase matching of the fiber taper guided mode to the whispering-gallery modes of
the microdisk\cite{ref:Spillane2} critical coupling was possible with a loaded $Q \sim 10^6$.  The fiber taper and
microdisk are then permanently attached using UV curable epoxy in two regions: (i) microscopic glue joints between the
fiber taper and lithographically defined SiN$_x$ supports (see Fig.\ 1(a)) fix the position of the taper relative to the
disk, and (ii) macroscopic glue joints between the taper support slide and the atom chip (see Fig.\ 1(d)) fix the
position of the taper support relative to the chip and serve as stress relief points for the fiber pigtails.  To avoid
blocking trapping laser beams or obscuring imaging, the entire fiber taper mount must lie below the plane of the
optically and magnetically trapped atoms ($\sim 600$ $\mu$m above the atom chip surface).  A sufficiently low-profile is
achieved by aligning and bonding the taper support glass coverslip parallel to and below the plane of the microdisk chip
top surface.  During the taper mounting procedure the taper-microdisk coupling is monitored by measuring $\lambda_o$ and
the transmission contrast ($\Delta T$), with no noticable change being observed during the curing of the epoxy joints.
The micro glue joints incur a taper diameter dependent amount of broadband insertion loss; for the taper diameters used
here ($\sim 1$ $\mu$m) an approximate $10$-$15\%$ optical loss per joint is typical.  Post-cure, the fiber-cavity
alignment is extremely robust, withstanding all of the vacuum installation procedures described below.

\begin{figure}[t]
\begin{center}
  \epsfig{figure=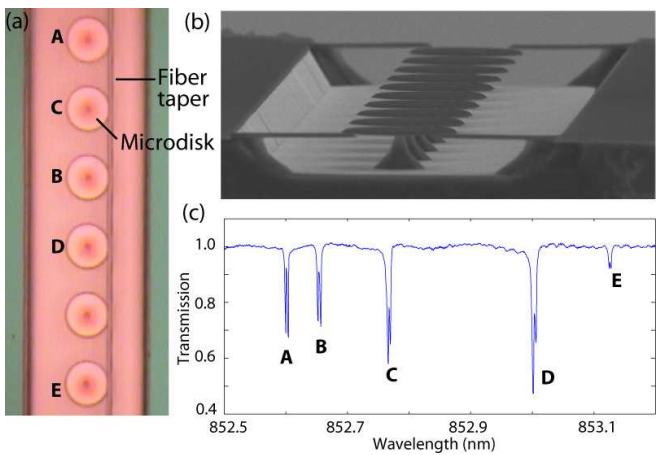, width=1.0\linewidth}
  \caption{(a) Top-view optical image of a fiber taper aligned with and array of 10 microdisks (the remaining 4
microdisks are out of the field of view of the image). (b) SEM image of the array of 10 microdisks. (c) Fiber taper
transmission vs.\ wavelength when the taper is aligned with the microdisk array in (a).  Letters match specific
microdisks in (a) with the corresponding resonances in (c).}
\label{fig:mult_disk}
\end{center}
\end{figure}

The integrated atom-cavity chip was installed in an ultra-high vacuum (UHV) chamber designed for performing atom chip
waveguiding experiments.  Vacuum-safe fiber feedthroughs\cite{ref:Abraham1} were used to pass the fiber-pigtails out of
the chamber.  The chamber was evacuated using turbo and ion pumps and baked at 130 $^\text{o}$C for $24$ hours so that a
background pressure of $< 10^{-8}$ Torr was reached.  Initial experiments with the integrated system involved the
trapping of Cs atoms in a mirror-MOT\cite{ref:Reichel} above the atom chip gold mirror and transfer of atoms to a
micro-U-MOT \cite{ref:Reichel} for subsequent magnetic trapping and guiding on the atom chip (see Fig.\ 1(d)).  The
extremely low-profile of the fiber taper mounting was shown to provide excellent optical access for atom trapping,
cooling, and imaging.  The microdisk resonance was continuously monitored during these procedures.  $\Delta T$ was found
to remain constant during the chamber pump-down, bake-out, and atom trapping, demonstrating the robustness of the
fiber-cavity mounting.  However, after the chamber bake, $\lambda_o$ was red shifted $0.11$ nm from the pre-bake value
and was found to increase logarithmically with exposure time to the Cs vapor as shown in Fig.\ 2(d).  During subsequent
chamber bakes and intermittent closures of the Cs source (for periods as long as 2 weeks) $\lambda_o$ was found to
remain constant.

The logarithmic time dependence of $\lambda_o$ with Cs exposure suggests that the Cs coverage of the microdisk surface
is saturating in a ``glassy'' manner \cite{ref:Anderson1}; interactions between deposited atoms quench the rate of
adsorption.  A shift $\Delta\lambda_o$ of the disk resonances can be related to a surface film thickness by $s \sim
\Delta\lambda_o /( \lambda_o (n_{\text{f}} - 1) \Gamma' )$, where $\Gamma' s$ represents the fraction of modal energy in
the film and $n_{\text{f}}$ is the refractive index of the film.  From finite element simulations of the microdisk,
$\Gamma' = 0.0026$ nm$^{-1}$ for the $p=1$ TE-like mode.  Assuming a film index of refraction equal to that of SiN$_x$,
the measured wavelength shift at the longest measured time ($t=450$ h) corresponds to roughly a half-monolayer coverage
of Cs on the disk surface (monolayer thickness $\sim 4\AA$ \cite{ref:Brause}).

As a future method of compensating for resonant detuning of the microdisk mode due to variation in fabrication or the
time-dependent Cs surface coverage, and as an initial demonstration of the scalability of the fiber-coupled microdisk
chip concept, we show in Figure 3(a) a single fiber taper coupled in parallel to an array of 10 nominally identical
microdisks (Fig.\ 3(b)).  Over a $\pm 0.25$ nm wavelength range the fiber taper couples to 5 of the $p=1$, TE-like modes
as shown in Fig.\ 3(c), the remaining 5 modes lying within a $\pm 1$ nm range. A modest 10 $^\text{o}$C of temperature
tuning may be used to tune between the 5 closely spaced modes.  Each of these resonances is due to coupling to a unique
microdisk, as verified by imaging the scattered light from the microdisks as a function of wavelength.

In conclusion, we have shown that wavelength-scale high-$Q$ microcavities can be realized from SiN$_x$ at near-visible
wavelengths, and have demonstrated a method for integrating these devices with atom chips.  The resulting optical fiber
taper interface to the hybrid atom-cavity chip provides sufficient optical access for chip-based atom trapping and
cooling while providing highly efficient optical coupling to single, or simultaneously to multiple, microdisk cavities.
In the future, the use of SiN$_x$ microcavities provides a path to a fully monolithic atom-cavity Si chip.

The authors thank M. Borselli and T. Johnson for assistance in FEM simulations and microdisk process development, and J.
Kerckhoff for help with Cs adsorbtion measurements.

\end{document}